\documentclass[aps,prev,twocolumn,preprintnumbers,floatfix,nofootinbib]{revtex4-1}
\usepackage{graphicx}
\usepackage{bm}
\usepackage{times}
\usepackage{slashed}
\usepackage{color}
\usepackage{appendix}
\usepackage{mathtools}
\usepackage{epsfig}
\usepackage{dcolumn}
\usepackage{textcomp}
\usepackage{multirow}

\begin{document} 

%\preprint{IIPDM-2021}

\title{A 2HDM for the g-2 and Dark Matter}

\author{Giorgio Arcadi$^{1}$}
\author{\'Alvaro S. de Jesus$^{2,3}$}
\email{alvarosdj@ufrn.edu.br}
\author{T\'essio B. de Melo$^2$}
\email{tessiomelo@gmail.com}
\author{Farinaldo S. Queiroz$^{2,3}$}
\author{Yoxara S. Villamizar$^{2,3}$}
\email{yoxara@ufrn.edu.br}
\affiliation{$^1$ Dipartimento di Matematica e Fisica, Universita di Roma 3, Via della Vasca Navale 84, 00146, Roma, Italy}
\affiliation{$^2$ International Institute of Physics, Universidade Federal do Rio Grande do Norte, Campus Universitario, Lagoa Nova, Natal-RN 59078-970, Brazil}
\affiliation{$^3$ Departamento de F\'isica, Universidade Federal do Rio Grande do Norte, 59078-970, Natal,
RN, Brasil}

\begin{abstract}

\noindent

The Muon g-2 experiment at FERMILAB has confirmed the muon anomalous magnetic moment anomaly, with an error bar 15\% smaller and a different central value compared with the previous Brookhaven result. The combined results from FERMILAB and Brookhaven show a difference with theory at a significance of $4.2\sigma$, strongly indicating the presence of new physics. In light of this new result, we  discuss a Two Higgs Doublet model augmented by an Abelian gauge symmetry that can simultaneously accommodate a light dark matter candidate and $(g-2) _\mu$, in agreement with existing bounds.
\end{abstract}


\maketitle
\flushbottom

\section{\label{Intro} Introduction}

\begin{figure}[!t]
    \centering
    \includegraphics[scale=0.43]{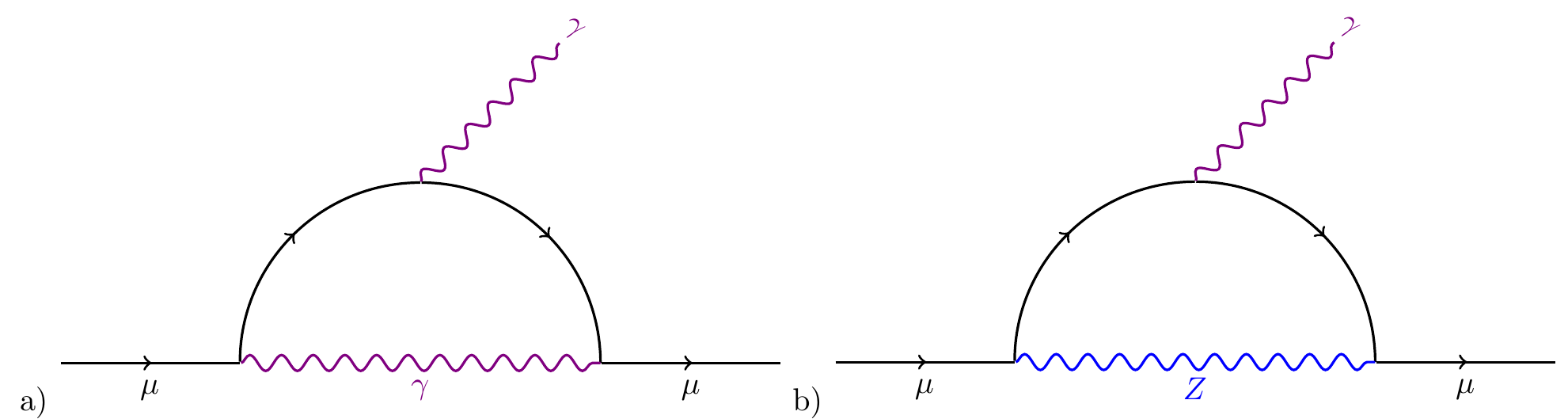}
    \includegraphics[scale=0.43]{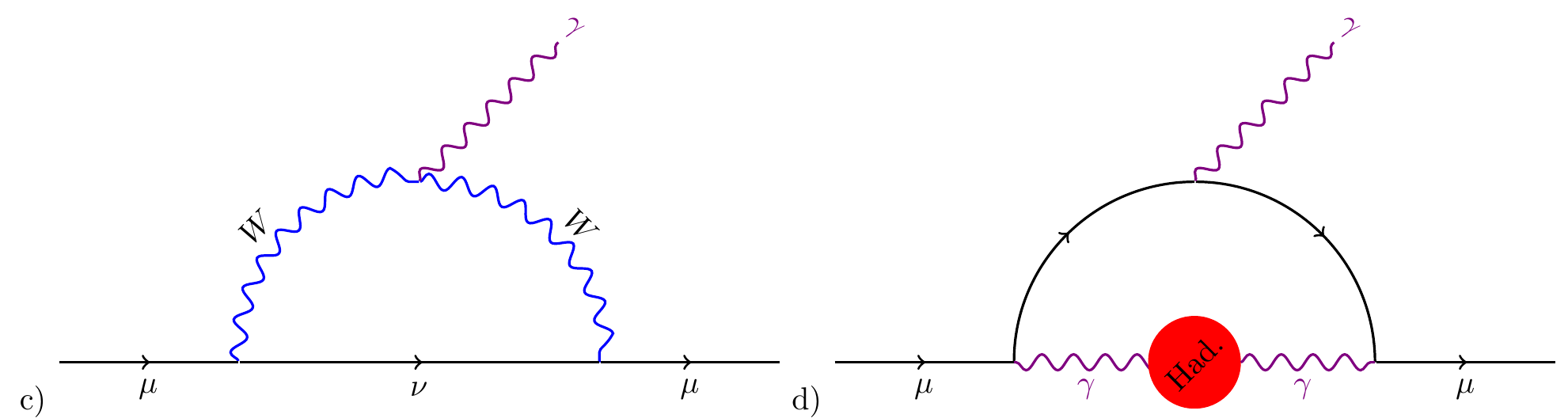}
    \caption{Feynman diagrams of the corrections to $a_{\mu}$ on SM interactions: a) first order QED, b) and c) lowest-order weak, and d) lowest-order hadronic effects.}
    \label{g2mu}
\end{figure}

The anomalous magnetic moment is a striking example of the extraordinary success of quantum field theory in explaining the fundamental properties of the elementary particles. According to the Dirac's theory, the magnetic dipole moment $\vec{\mu}$ of a fermion of mass $m$, electric charge $q$, and spin $\vec{s}$ is,
\begin{equation}
\vec{\mu} = g \frac{q}{2 m} \vec{s} , 
\end{equation}
with $g$ being the gyromagnetic ratio, predicted to be $g = 2$. Quantum Electrodynamics predicts departures at higher orders in perturbation theory, from this exact result. Such departures are parametrized by the anomalous magnetic moment $a= \frac{g-2}{2}$. Further contributions to the anomalous magnetic moment, described by the Feymann diagrams in Fig. \ref{g2mu}, come from weak and strong interactions of the Standard Model (SM). New Physics can be, as well, responsible for radiative corrections to the anomalous magnetic moment. The latter could be then constrained or possibly discovered by looking for eventual discrepancies between the SM prediction for $a$ and its corresponding experimental measure. 

This is the case of the muon anomalous magnetic moment $a _\mu$. There is indeed a long-standing discrepancy between the SM theoretical prediction \cite{Aoyama:2020ynm},
\begin{equation}
\label{sm_pred_a_mu}
a _\mu ^{\text{SM}} = 116 591 810(43) \times 10 ^{-11} , 
\end{equation}
and the experimental measurements, which has driven a multitude of studies in the past decades (see Ref. \cite{Lindner:2016bgg} for a review). 
The discrepancy $\Delta a _\mu = a _\mu ^{\text{exp}} - a _\mu ^{\text{SM}}$ depends on hadronic light-by-light and vacuum polarization  \cite{Bennett:2002jb,Bennett:2006fi}, whose computation rely on dispersion relations and experimental input data from $e ^+ e ^- \rightarrow$ hadrons cross section, which is subject to large uncertainties. Consequently, the significance of the anomaly has varied over the years,

\begin{eqnarray}
    \Delta a_\mu & = & (261 \pm 78)\times 10^{-11}\,\, (3.3\sigma)\,\, \text{ \cite{Prades:2009tw,Tanabashi:2018oca} - (2009)}; \nonumber\\
     \Delta a_\mu & = &  (325 \pm 80)\times 10^{-11}\,\, (4.05\sigma)\,\, \text{ \cite{Benayoun:2012wc} - (2012)};\nonumber\\
     \Delta a_\mu & = & (287 \pm 80)\times 10^{-11}\,\, (3.6\sigma)\,\, \text{ \cite{Blum:2013xva} - (2013)};\nonumber\\
     \Delta a_\mu & = &  (377 \pm 75)\times 10^{-11}\,\, (5.02\sigma)\,\, \text{ \cite{Benayoun:2015gxa} - (2015)};\nonumber\\
          \Delta a_\mu & = &  (313 \pm 77)\times 10^{-11}\,\, (4.1\sigma)\,\, \text{ \cite{Jegerlehner:2017lbd}- (2017)};\nonumber\\
     \Delta a_\mu & = &  (270 \pm 36)\times 10^{-11}\,\, (3.7\sigma)\,\, \text{ \cite{Keshavarzi:2018mgv} - (2018)};
     \label{deltasigmavalues}
\end{eqnarray}
depending on the different values adopted for the hadronic contributions.

On recent times, an exiting new measurement has been announced by the Muon g-2 Experiment at FERMILAB \cite{Abi:2021gix,Albahri:2021kmg,Albahri:2021ixb}, exceeding again the SM prediction by around $4.2 \sigma$,
\begin{equation}
a _\mu ^{\text{exp}} = 116 592 061(41) \times 10^{-11},
\end{equation}
leading to an update on the value of $\Delta a _\mu$,
\begin{equation}
\label{delta_a_mu_disc}
\Delta a _\mu = 251(59) \times 10^{-11} .
\end{equation}
Although $\Delta a _\mu$ may reach a $5 \sigma$ significance in the near future with improved experimental precision, a recent lattice simulation of the hadronic contribution from the BMW group \cite{Borsanyi:2020mff} suggests a modified value for the SM prediction, which is much closer to the value observed in the FERMILAB experiment. If correct, the BMW result would reduce the significance to the $1.5 \sigma$ level. This result, although more precise than the previous lattice calulations, differs from the ones obtained by the data-driven approach and leads to tensions with global electroweak fits, via the change in the hadronic running of the fine-structure constant \cite{Crivellin:2020zul}. At the moment, the situation is quite unclear and further theoretical and experimental efforts will be required to settle this question.
In this work we will adopt the value given in Eq. (\ref{sm_pred_a_mu}), based on the data-driven approach, as the reference SM value for $a _\mu$, which leads to the $\Delta a _\mu$ discrepancy given in Eq. (\ref{delta_a_mu_disc}).

%This observation may reach a $5 \sigma$ significance in the near future with refined experimental precision and improved theoretical calculations.
%Nevertheless, a recent lattice simulation of the hadronic contribution from the BMW group \cite{Borsanyi:2020mff} suggests a modified value for the SM prediction, which is close to the one observed in the FERMILAB experiment, reducing the significance to $1.5 \sigma$. This result, although much more precise than the previous lattice calulations, differs from the ones obtained by the data-driven approach and leads to tensions with global electroweak fits, via the change in the hadronic running of the fine-structure constant \cite{Crivellin:2020zul}. At the moment, the situation is still unclear and further theoretical and experimental efforts will be required to settle this question.
%In this work we will adopt the value given in Eq. (\ref{sm_pred_a_mu}), based on the data-driven approach, as the reference SM value for $a _\mu$, which leads to the $\Delta a _\mu$ discrepancy given in Eq. (\ref{delta_a_mu_disc}).

Interpreting such a discrepancy as an imprint of New Physics \cite{Athron:2021iuf, Das:2021zea, Das:2020hpd, Greljo:2021xmg, Jana:2020pxx, Jana:2020joi, Cvetic:2020vkk, Keus:2017ioh}, here we discuss a Two Higgs Doublet Model (2HDM) augmented by an Abelian $U(1) _{L _\mu - L _\tau}$ gauge symmetry.  
The $L _\mu - L _\tau$ model is an economical SM extension, since it is well-known that the $U(1) _{L _\mu - L _\tau}$ symmetry is anomaly free without the need of extra fermions. This model has been studied in the context of the $(g-2) _\mu$ anomaly \cite{Baek:2001kca, Gninenko:2018tlp, Ma:2001md, Harigaya:2013twa}, since the new $Z ^\prime$ gauge boson, coupled directly to the muon, provides the required contribution to its magnetic moment, while avoiding strong bounds stemming from low energy electron probes. The $L _\mu - L _\tau$ model has been investigated also in several other contexts, such as in B meson decays \cite{Crivellin:2015mga,Altmannshofer:2016jzy}, Higgs flavor violating decays \cite{Crivellin:2015mga, Lee:2014rba}, LFV processes \cite{Heeck:2011wj, Dutta:1994dx}, muon colliders \cite{Huang:2021nkl,Huang:2021biu}, dark matter \cite{Arcadi:2018tly,Foldenauer:2018zrz, Biswas:2016yjr} and neutrino masses \cite{Heeck:2014qea, Biswas:2016yan}.

Differently from the usual $L _\mu - L _\tau$ model in which the $U(1) _{L _\mu - L _\tau}$ is typically broken by a scalar $SU(2)$ singlet, here we consider the case in which the symmetry is spontaneously broken by a scalar $SU(2)$ doublet. In such a case, the mass of the $U(1) _{L _\mu - L _\tau}$ ($Z^{\prime}$) gauge boson is predicted to be below the electroweak scale. Furthermore, mass mixing between the $Z^{\prime}$ and the SM $Z$-boson is expected. This latter feature opens further opportunity to probe the model via parity violation experiments. 
Besides discussing a 2HDM model with an $L_\mu-L_\tau$ symmetry in connection with the $(g-2) _\mu$, we present a simple addition of a successful light dark matter candidate in agreement with existing limits.

Our paper is structured as follows: In section II we introduce the model and discuss its key aspects; In section III we review the relevant constraints and display the region of parameter space that accounts for the $(g-2) _\mu$ anomaly, highlighting the differences from the scenario in which the new gauge symmetry is broken by a $SU(2)$ singlet.
In the section IV we add a dark matter candidate and explain how it can reproduce the correct relic density and the $(g-2) _\mu$ anomaly at the same time; Section V is devoted to discussion of the results and further comparison with similar models. We finally state our conclusions in section VI.

\section{\label{model}Model}

The model we are going to study in this work extends the SM with a gauged $L _\mu - L _\tau$ symmetry. This is a straightforward extension of the SM global $U(1) _{L _\mu - L _\tau}$ symmetry, being the latter anomaly free. 
Under this symmetry, the fermion charges are set by the difference between their muon and tau lepton numbers; hence only the second and third SM lepton generation have not null charge under this symmetry (see Tab. \ref{charges_u1_mu_tau} for a summary). 
Therefore, the $L _\mu - L _\tau$ gauge boson $Z ^\prime$ couples directly only to the muon, tau and respective neutrinos, a feature that is desirable in order to address the $(g-2) _\mu$ anomaly, since it allows the $Z ^\prime$ to give a sizeable contribution to the muon magnetic moment, while avoiding the bounds from several experimental probes, as we will discuss in more detail below.

The lepton gauge interaction reads,
\begin{equation}
\label{LagZ}
- \mathcal{L} _{Z ^\prime f f} = \frac{g ^\prime}{2} (\bar{\mu} \gamma^{\mu} \mu + \bar{\nu}_{\mu} \gamma^{\mu} P_{L} \nu_{\mu} - \bar{\tau} \gamma^{\mu} \tau - \bar{\nu}_{\tau} \gamma^{\mu} P_{L} \nu_{\tau} ) Z_{\mu}^{\prime} ,
\end{equation}
where $g ^\prime$ is the gauge coupling constant and $P _L$ is the left chiral projector. 
This lepton non-universal coupling to muon, tau and neutrinos is the primary interaction mode of $Z ^\prime$ with the fermions. Despite being uncharged under $U(1) _{L _\mu - L _\tau}$, 
the other fermions interact with $Z ^\prime$ via kinetic mixing,
\begin{equation}
\label{int_kin_mix}
\mathcal{L} _{{Z ^\prime} _{kin}} = - \epsilon e J _{em} ^\mu Z _\mu ^\prime ,
\end{equation}
which arises from the mixing of the field strengths of $U(1) _{L _\mu - L _\tau}$ and $U(1) _Y$ via the renormalizable operator $\frac{\epsilon}{2} F ^{\mu \nu} F _{\mu \nu} ^\prime$,
where $\epsilon$ is the kinetic mixing parameter. Even if $\epsilon$ is set to zero at tree level, quantum corrections involving muons and taus running in the loops result in the mixing of the photon and $Z ^\prime$ propagators, leading to a nonzero $\epsilon$. 
At 1-loop level, it is given by \cite{Araki:2017wyg},
\begin{equation}
\label{epsilon}
\epsilon = \frac{4 e g ^\prime}{(4 \pi) ^2} \int _0 ^1 x (1 - x) \ln \frac{m _\tau ^2 - q ^2 x (1 - x)}{m _\mu ^2 - q ^2 x (1 - x)} dx .
\end{equation}
We assume that the kinetic mixing is set by this finite calculable quantity, which means that, differently from dark photon models \cite{Holdom:1985ag, Pospelov:2007mp, Arkani-Hamed:2008hhe} in which $\epsilon$ is typically considered to be a constant free parameter, here it will be fixed in terms of $g ^\prime$.
The dependence on the momentum transfer $q$ is such that $\epsilon$ is suppressed by $1 / q ^2$ for large momentum $q ^2 \gg m _\tau ^2$, while it becomes approximately constant $\epsilon \sim 10 ^{-2} g ^\prime$ for small $q ^2$. In general, the loop-induced $\epsilon$ will be smaller than $g ^\prime$ typically by at least two orders of magnitude, 
which makes the interaction in Eq. (\ref{int_kin_mix}) quite feeble. Despite being suppressed by the small $\epsilon$, the $Z ^\prime$ coupling to the electron leads to measurable effects on neutrino-electron scattering experiments and electron-positron colliders. As for the muon and tau leptons, the kinetic mixing can often be neglected in face of the coupling in Eq. (\ref{LagZ}). In particular, the kinetic mixing yields negligible contribution to the muon magnetic moment.

\begin{table}[!t]
\begin{tabular}{|c|c|c|c|c|c|c|c|c|c|c|c|}
\hline 
\multirow{2}{*}{Fields} & \multicolumn{2}{c|}{Quarks} & \multicolumn{6}{c|}{Leptons} & \multicolumn{2}{c|}{Scalars} \\ \cline{2-11}
 & $u _i$ & $d _i$ & $e$ & $\mu$ & $\tau$ & $\nu _e$ & $\nu _\mu$ & $\nu _\tau$ & $\Phi _1$  & $\Phi _2$ \\ \hline 
$U(1) _{\mu-\tau}$ Charges & 0 & 0 & $0$ & $1$ & $-1$ & $0$ & $1$ & $-1$ & $Q _1$ & $0$  \\ \hline 
\end{tabular}
\label{charges_u1_mu_tau}
\caption{{\footnotesize $U(1)_{\mu - \tau}$ charge assignment for leptons and scalars. Only the second and third lepton families are charged under $U(1)_{\mu - \tau}$, and all quarks are neutral. The charge of $\Phi _1$ is assumed to be nonzero, $Q _1 \neq 0 $.}}
\end{table}

\begin{figure}[!t]
	\centering
	\includegraphics[width=1\columnwidth]{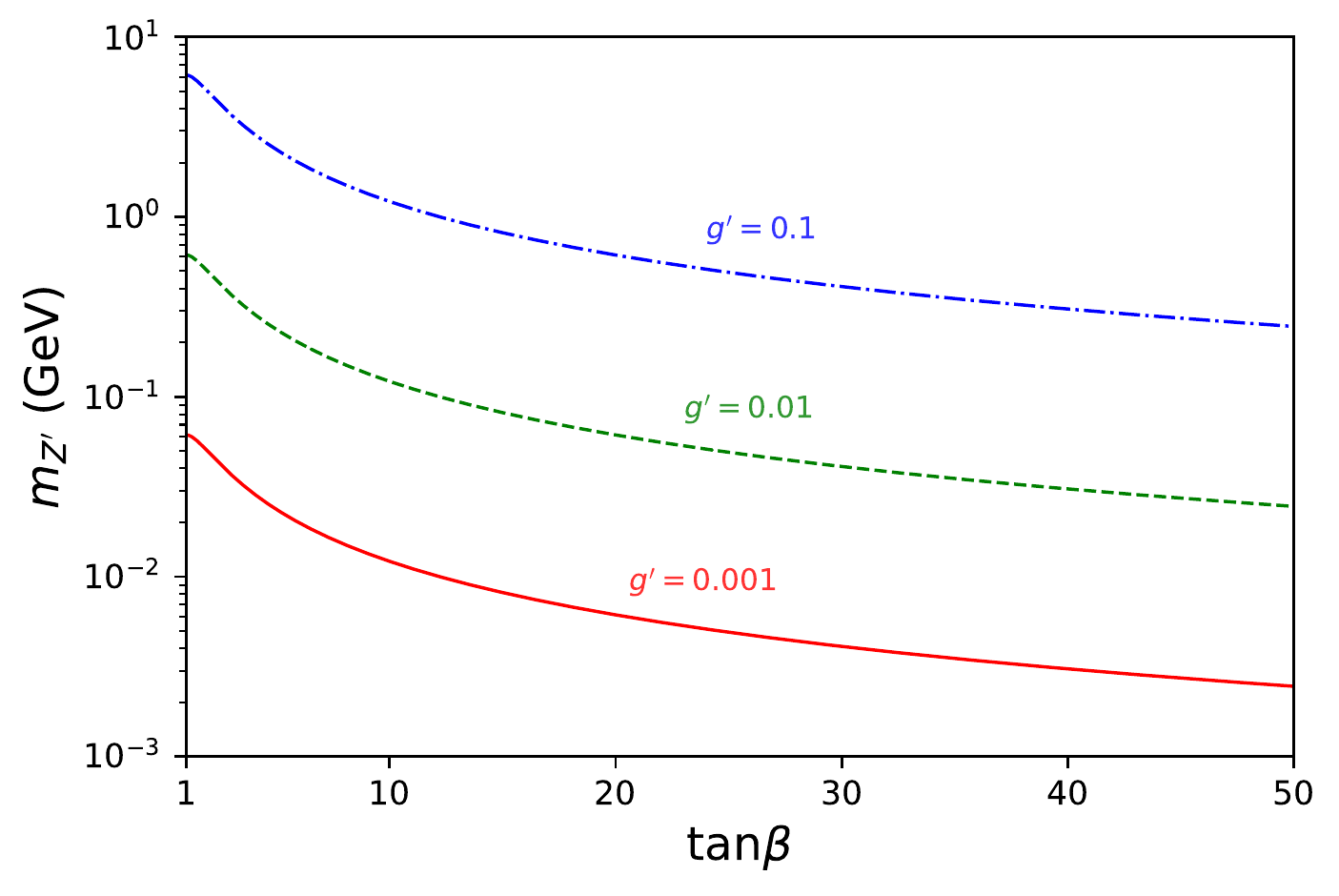}
	\caption{$Z^\prime$ mass as a function of $\tan \beta$ and gauge coupling constant $g ^\prime = 10 ^{-2}, 10 ^{-3}, 10 ^{-4}$ (the $\Phi _1$ charge is fixed as $Q _1 = 1$). For $g ^\prime < 10 ^{-2}$, the $Z ^\prime$ is in the sub-GeV range for any value of $\tan \beta$.}
	\label{zp_mass}
\end{figure}

In order to break the $U(1) _{L _\mu - L _\tau}$ symmetry, an extra $SU(2) _L$ scalar doublet is introduced, resulting in a scalar sector containing two doublets,
\begin{equation*}
\Phi _i = \begin{pmatrix} \phi _i ^+ \\ \phi _i ^0 \end{pmatrix} \text{\ \ \ \ \ \ } \sim \text{\ \ \ \ \ \ } ( 1, 2, 1, Q _i ) ,
\end{equation*}
where $i = 1, 2$ and the charges above correspond to the quantum numbers under the $SU(3) _c \times SU(2) _L \times U(1) _Y \times U(1) _{L _\mu - L _\tau}$ group, respectively.
We assume that the $\Phi _2$ doublet is neutral under $U(1) _{L _\mu - L _\tau}$ and that the charge of $\Phi _1$ is a nonzero free parameter, $Q _1 \neq 0 $. The vacuum expectation value (VEV) of $\Phi _1$ triggers the spontaneous breaking of $U(1) _{L _\mu - L _\tau}$, generating the mass for the new gauge boson dynamically. Since the doublet VEVs must satisfy the condition $(v _1 ^2 + v _2 ^2) ^{1/2} \equiv v = 246$ GeV, the $U(1) _{L _\mu - L _\tau}$ symmetry will be broken below the electroweak scale, which constrains the $Z ^\prime$ mass towards lower values (in contrast with the case in which the symmetry is broken by a scalar singlet, whose VEV has no restrictions \textit{a priori}).
After the spontaneous symmetry breaking, the $Z ^\prime$ acquires a mass given by\footnote{Here we implicitly have to take the absolute value of $Q _1$, so that $m _{Z ^\prime}$ is always positive.},
\begin{equation}
\label{zp_mass_eq}
m _{Z ^\prime} = \frac{1}{2} g ^\prime Q _1 v \sin \beta \cos \beta ,
\end{equation} 
where the angle $\beta$ is defined from the ratio of the doublet VEVs, $\tan \beta = v _2 / v _1$. From the trigonometric dependence on $\beta$, it is clear that $m _{Z ^\prime}$ has a maximum value at $\tan \beta = 1$, for fixed $g ^\prime$ and $Q _1$ (see Fig. \ref{zp_mass}). We see that $m _{Z ^\prime}$ naturally lies below the electroweak scale and becomes sub-GeV for $g ^\prime \lesssim 0.1$. As we will see in the next section, the parameter space favored by the $(g-2) _\mu$ anomaly requires $g ^\prime \sim O (10 ^{-3})$ and $m _{Z ^\prime}$ around $10-100$ MeV, which can be reached with mild $\tan \beta$ values.

Since the VEV of the $\Phi _1$ doublet breaks simultaneously the $U(1) _{L _\mu - L _\tau}$ and the electroweak symmetries, a mass mixing between the $Z ^\prime$ and the SM $Z$ boson will be induced, resulting in a $Z ^\prime$ coupling to the weak neutral current $J _{NC} ^\mu$, 
\begin{equation}
\mathcal{L} _{Z ^\prime mix} = - \epsilon _Z g _Z J _{NC} ^\mu Z _\mu ^\prime ,
\label{zp_mass_mixing}
\end{equation}
where $g _Z = g / \cos \theta _W$. The mass mixing parameter $\epsilon _Z$ represents essentially the off-diagonal entry in the $Z-Z ^\prime$ mass matrix, and is determined by the charge of the $\Phi _1$ doublet and by the $g ^\prime$ and $\beta$ parameters, according to $\epsilon _Z g _Z = g ^\prime Q _1 \cos ^2 \beta$. Notice that $\epsilon _Z$ is suppressed in the regime $\tan \beta > 1$, which we assume here. 
This additional suppression renders the interaction (\ref{zp_mass_mixing}) negligible for most purposes, except by the fact that it introduces parity violation through the axial $Z ^\prime$ couplings to the fermions, which can be probed in scattering experiments with polarized electrons and parity violating transitions in atomic systems \cite{Campos:2017dgc, Arcadi:2019uif,Davoudiasl:2012qa, Davoudiasl:2012ag}. This parity violation induced by mass mixing is a distinct feature for the case in which the symmetry is broken by a scalar doublet, in comparison to the case when it is broken by a singlet.

The masses of SM quarks and leptons are generated by the $U(1) _{L _\mu - L _\tau}$ neutral doublet $\Phi _2$ through the Yukawa Lagrangian,
\begin{equation}
- \mathcal{L} _Y = y ^d \bar{Q} \Phi _2 d _R + y ^u \bar{Q} \widetilde \Phi _2 u _R + y ^e \bar{L} \Phi _2 e _R + h.c.
\end{equation}
where $Q$ ($L$) is the left-handed quark (lepton) doublet, $u_R$ ($d_R$) is the right-handed up (down) quark and $e_R$ is the right-handed charged lepton. As long as $Q _1 \neq Q _2$ the coupling of $\Phi _1$ to the fermions is forbidden by the $L _\mu - L _\tau$ symmetry\footnote{It is possible to have a term in the Yukawa Lagrangian such as $ y ^{\mu \tau} \bar{L} ^\tau \Phi _1 \mu _R $, connecting the second and third lepton families via the doublet $\Phi _1$. This term gives rise to off-diagonal elements in the lepton mass matrix, generating mixing in the lepton sector. The consequences of this mixing for anomalies in B meson decays and Higgs flavor violating decays have been studied, e.g., in Refs. \cite{Crivellin:2015mga, Heeck:2014qea}. In this work we will not consider these off-diagonal lepton interactions.}, preventing the appearance of Higgs mediated flavor changing neutral currents, a well known issue in the context of 2HDMs. We assume that the condition $Q _1 \neq Q _2$ is always satisfied here, since $Q _1 \neq 0$ is necessary in order that $\Phi _1$ break the $U(1) _{L _\mu - L _\tau}$ symmetry.  
Therefore the model is free from tree level flavor changing neutral currents and the fermions couple to the scalars as in a standard type-I 2HDM \cite{Branco:2011iw}.

\begin{figure}[!t]
\centering
\includegraphics[width=0.63\columnwidth]{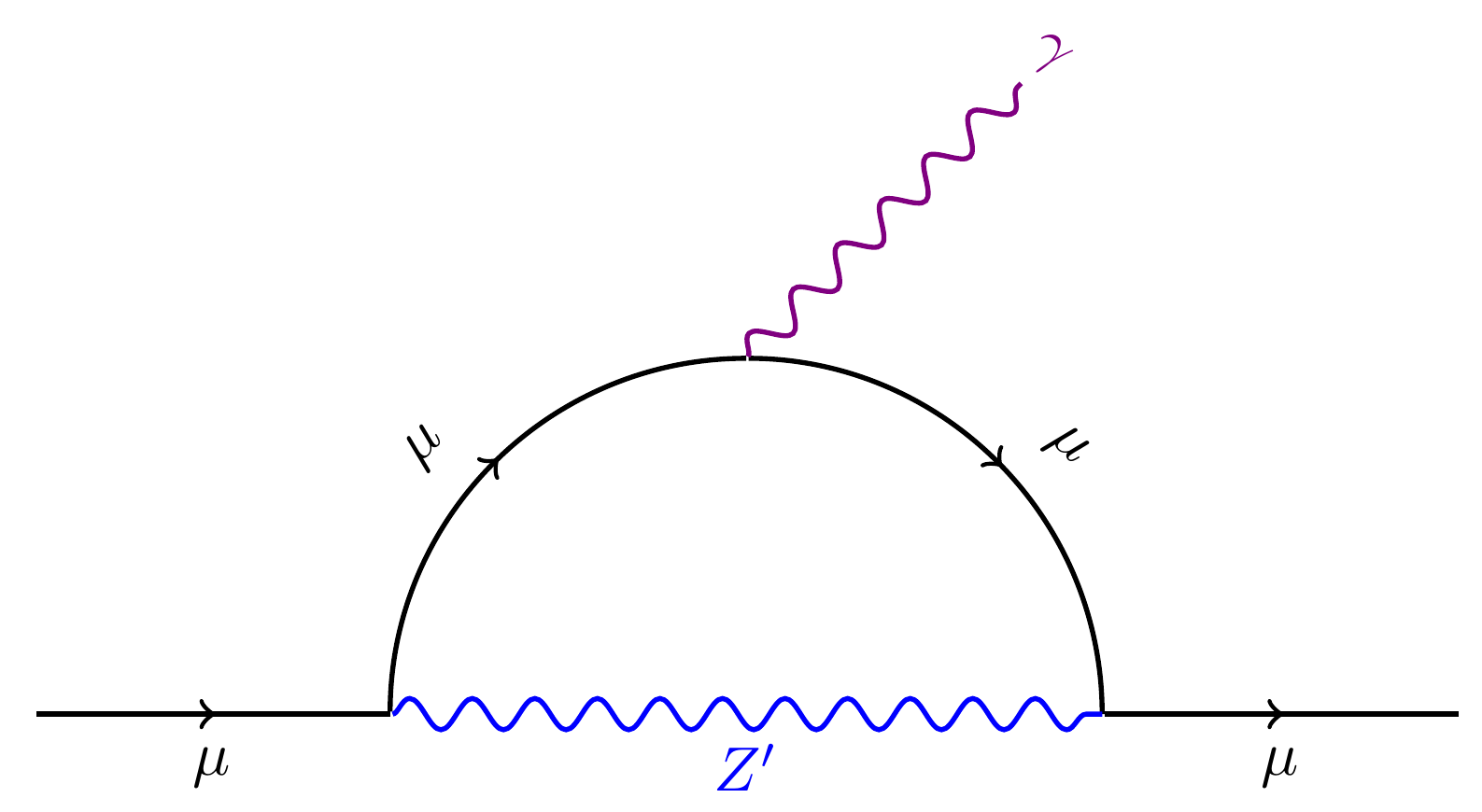} 
\caption{One-loop Feynman diagram of the $Z^\prime$ contribution to the muon anomalous magnetic moment.}
\label{loopdiagram}
\end{figure}

The scalar sector features four particles: the two neutral bosons $h$, $H$ and the charged pair $H ^\pm$, where the CP-even scalar $h$ is identified as the 125 GeV state found at the LHC. Differently from the usual 2HDM, this model does not contain a physical pseudoscalar, which becomes the longitudinal component of the $Z ^\prime$.
By and large, the extra scalars $H$ and $H ^\pm$ are allowed to be either heavier or lighter than the SM-like Higgs $h$. However, barring fine tunning in the potential parameters, $H$ cannot be much heavier than $125$ GeV, since its mass is roughly proportional to $\cot \beta$. 
The mass of $H ^+$ does not depend on $\tan \beta$, and thus can be much larger than $m _H$, being limited only by the perturbativity of the quartic potential couplings. However, electroweak precision data tends to constrain large hierarchies in the scalar masses, so that $m _{H ^+}$ should track $m _H$ to some extent, and we can expect a relatively light scalar sector.

\begin{figure}[!t]
  \centering
  \includegraphics[width=\linewidth]{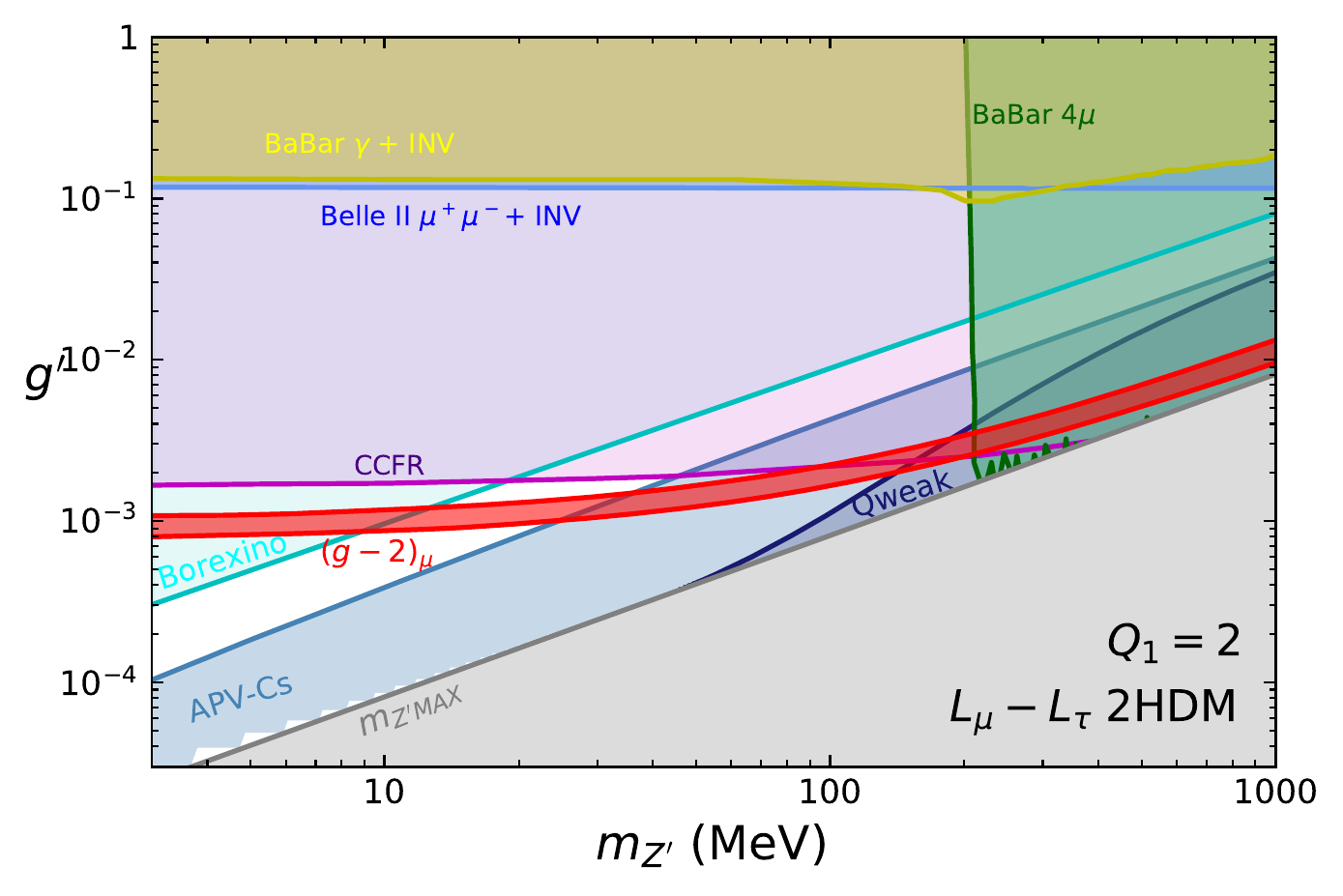}
  \caption{Allowed parameter space on the $Z^\prime$ mass and gauge coupling constant $g^\prime$ to explain the $(g-2) _\mu$ anomaly and exclusion regions from different experiments: $e^+e^-$ colliders BaBar (yellow) and Belle II (blue); neutrino-electron scattering from Borexino (cyan); neutrino trident production from CCFR (purple) and parity violation in Cesium (light blue) and proton (dark blue). The charge of the $\Phi _1$ doublet is fixed as $Q _1 = 2$. The $m _{Z ^\prime \text{\tiny{MAX}}}$ line indicates the maximum $m _{Z ^\prime}$ value (at $\tan \beta = 1$) for fixed $Q _1$, and the grey region below this line is forbidden.}
  \label{Plot}
\end{figure}

Contrary to other 2HDMs, in the type-I one can still have $H$ and $H ^+$ with masses as low as $\sim 100$ GeV without being in conflict with current data. The main constraints come from the direct searches for non standard Higgs bosons at the LHC \cite{ATLAS:2018gfm,CMS:2019bfg,ATLAS:2021upq,CMS:2019rlz,ATLAS:2020zms,CMS:2018rmh} and from flavor physics indirect searches, especially with B meson decays \cite{Haller:2018nnx, Misiak:2017bgg, Dumont:2014wha}. 
As in the type-I 2HDM all the couplings of $H$ and $H ^+$ to the fermions are proportional to $\cot \beta$, they become suppressed in the $\tan \beta > 1$ regime.
While, for instance, in type-II and type-Y 2HDMs, the measurement of the BR$(b \rightarrow s \gamma)$ constrains the mass of $H ^+$ to be larger than about $570$ GeV, in type-I this bound is evaded provided that $\tan \beta \gtrsim 2$.
Since in this regime the extra scalars give negligible contribution to the muon magnetic moment, they do not play a significant role in the $(g-2) _\mu$ phenomenology and, for this reason, we shall not extend the discussion about them any further. However, we would like to stress that the presence of these extra (possibly light) Higgs bosons is in agreement with current constraints. \\

Having presented the main features of the model, in the following we discuss the parameter space suitable to accommodate the $(g-2) _\mu$ anomaly and the constraints on that region coming from several low energy experiments. In the subsequent section, we discuss the extension of the model to include a dark matter sector.

\section{$(g-2) _\mu$ Parameter Space and Existing Constraints}

\begin{figure}[!t]
  \centering
  \includegraphics[width=\linewidth]{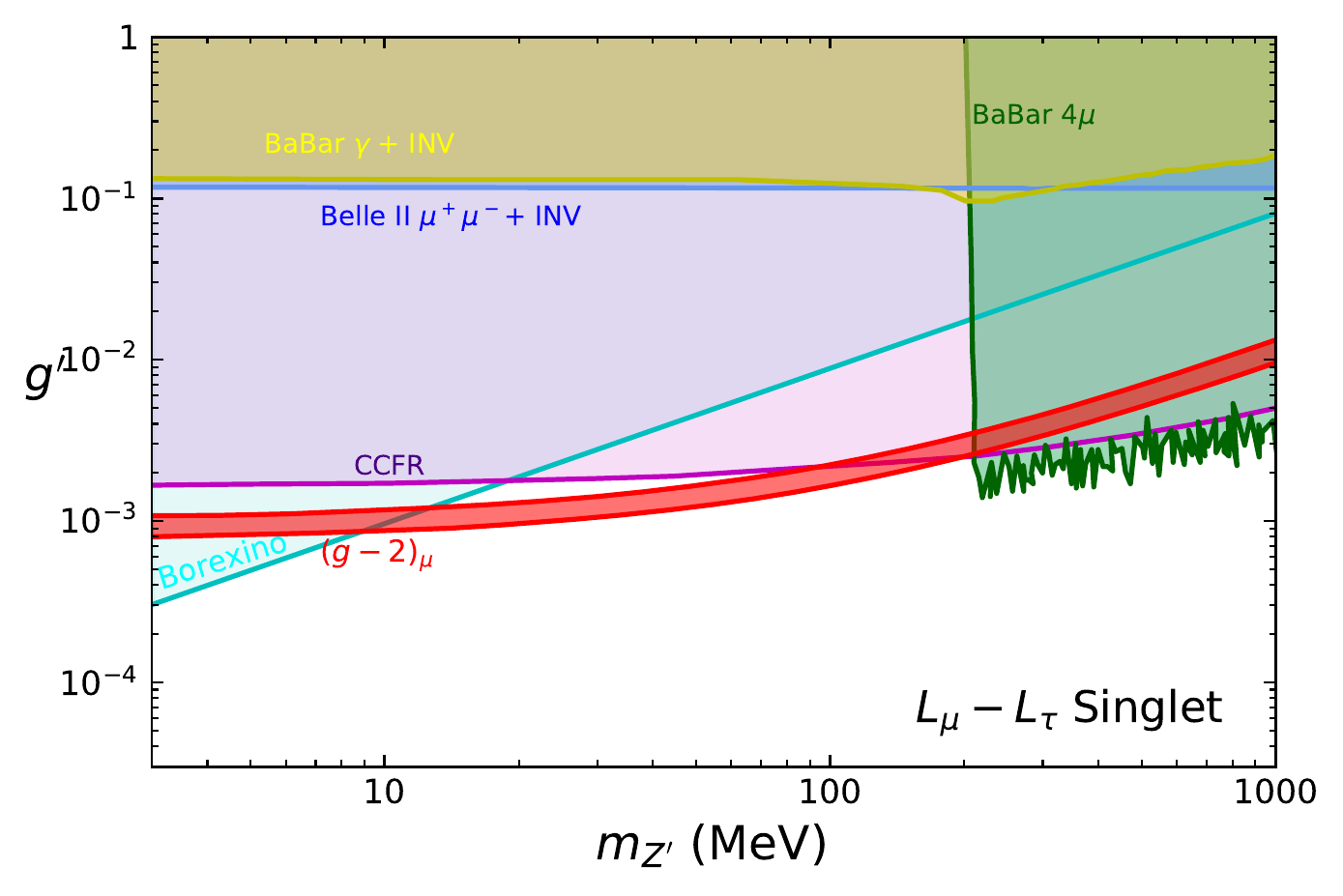}
  \caption{Similar plot as in Fig. \ref{Plot} for the case in which the $L _\mu - L _\tau$ symmetry is broken by a scalar singlet. In this case the bounds from parity violation no longer apply and the $Z ^\prime$ mass is unconstrained at higher values. The color code is the same used in the Fig. \ref{Plot}.}
  \label{PlotSing1}
\end{figure}

In this section we present the contribution to the muon anomalous magnetic moment via the new $Z ^\prime$ gauge boson and discuss the constraints on the parameter space of the $L _\mu - L _\tau$ model, which are summarized in the Fig. \ref{Plot}. 
The favoured $(g-2) _\mu$ region is represented by the red band in the plane $g ^\prime$ versus $m _{Z ^\prime}$. In order to illustrate the main differences with the scenario considered here, we also show in the Fig. \ref{PlotSing1} the corresponding bounds for the $L _\mu - L _\tau$ singlet setup.

In general, the bounds on the $L _\mu - L _\tau$ model tend to be weaker when compared to other $U(1)$ models with universal charges, since the particles of the first generation (which constitute most of the ordinary matter, including the materials inside the detectors) are neutral in the $L _\mu - L _\tau$ case. Therefore, experiments that probe directly the $Z ^\prime$ coupling to muons are expected to provide better sensitivities. Nevertheless, the indirect $Z ^\prime$ coupling to electrons via kinetic mixing with the photon play an important role in the phenomenology, leading to relevant bounds from experiments with elastic neutrino-electron scattering. In addition, the mass mixing with the $Z$ boson introduces new constraints from parity violation. Next we discuss each one in turn. 
\\

\paragraph{\label{gmu2}Muon anomalous magnetic moment}

The recent measurement from the Muon $g-2$ experiment at FERMILAB has strengthened the long standing $(g-2) _\mu$ anomaly, shrinking the error bar in 15\% and raising the statistical significance of the discrepancy with the SM prediction to the level of $4.2\sigma$. The contribution for $\Delta a _\mu$ from the $L _\mu - L _\tau$ model comes from the virtual exchange of the $Z ^\prime$ via the 1-loop diagram shown in Fig. \ref{loopdiagram}. 
Since the $Z ^\prime$ here features vector and axial couplings to the muon, the $\Delta a _\mu$ has two contributions,
\begin{equation}
\label{amuzp}
\Delta a _\mu = \frac{1}{8 \pi ^2} \frac{m _\mu ^2}{m _{Z ^\prime} ^2} \int _0 ^1 dx \frac{|g _V| ^2 P _4 ^+ (x) + |g _A| ^2 P _4 ^- (x)}{(1-x)(1 - \lambda ^2 x) + \lambda ^2 x} .
\end{equation}
In this expression, $g _V$ and $g _A$ are the vector and axial $Z ^\prime$ couplings to the muon (determined from the Eqs. (\ref{LagZ}), (\ref{int_kin_mix}) and (\ref{zp_mass_mixing})), $P _4 ^+ (x) = 2 x ^2 (1-x)$, $P _4 ^- (x) = 2 x (1-x) (x-4) - 4 x ^3 \lambda ^2$ and $\lambda = \frac{m _\mu}{m _{Z ^\prime}}$. 
Comparing the Eq. (\ref{amuzp}) with the reported experimental value $\Delta a_{\mu} = 251 (59) \times 10^{-11}$, one obtains the parameter space that accounts for the anomaly. The corresponding region in the $g ^\prime \times m _{Z ^\prime}$ plane within $1 \sigma$ is depicted as the red band in Fig. \ref{Plot}.

Although the contribution to $\Delta a _\mu$ from the axial part is negative, it turns out to be negligible compared to the right sign contribution from the vector current. In particular, we notice very little difference to the singlet case of Fig. \ref{PlotSing1}, in which the $Z ^\prime$ couplings are purely vectorial. \\

\paragraph{Neutrino-electron scattering}

The $Z^{\prime}$ gauge boson contributes to the elastic neutrino-electron scattering, although this contribution depends on the suppressed $Z^{\prime}$ coupling to the electron via $\gamma/Z^{\prime}$ mixing. Sizable constraints nonetheless apply. We show a cyan shaded region in Fig.\ref{Plot}, the exclusion region from the Borexino data on the scattering of low energy solar neutrinos \cite{PhysRevD.98.015005, Agostini:2017ixy}, being, at the moment, the strongest experimental exclusion for this type of process.
Compared to the bounds from other experimental probes, the Borexino limits become relevant especially in the low $m _{Z ^\prime}$, low $g ^\prime$ region. \\

\paragraph{Neutrino trident production}
Neutrino trident production is a powerful probe for new forces that couple to muons and muon neutrinos. It corresponds to the process $\nu N \rightarrow \nu N \mu^{+} \mu^{-}$ in which a muon neutrino is scattered off of a nucleus producing a $\mu ^+ \mu ^- $ pair. In the SM this process occurs via the $Z$ boson, and thus gets enhanced in the presence of $Z ^\prime$. 
The CCFR collaboration has detected neutrino trident events at levels consistent with the SM, which translates to strong bounds on the possible $Z ^\prime$ contributions, displayed in the Fig. \ref{Plot} in purple (adapted from Ref. \cite{PhysRevLett.113.091801}). \\

\paragraph{Searches in $e^- e^+$ colliders}

A light $Z ^\prime$ can be produced at low energy $e^- e^+$ colliders such as BaBar and Belle in a couple of different ways, e.g., associated with photons $e ^- e ^+ \rightarrow \gamma Z ^\prime $ via kinetic mixing, or associated with a muon pair $e^{+} e^{-} \rightarrow \mu^{+} \mu^{-} Z^{\prime}$ via the direct gauge coupling with muons. 
The mono-photon search at BaBar \cite{Lees:2017lec} looked for $Z ^\prime$ invisible decay events from the associated production with photons, which results in the yellow exclusion region in the Fig. \ref{Plot}. The BaBar collaboration also searched for events with $4 \mu$ in the final state \cite{TheBABAR:2016rlg}, from the associated production with a muon pair and the subsequent decay $Z ^\prime \rightarrow \mu \mu$, leading to constraints for $m _{Z ^\prime} > 2 m _\mu$, shown as the green region the Fig. \ref{Plot}. There are also bounds from the Belle II search for the $Z ^\prime$ associated production with muons in the invisible mode \cite{Adachi:2019otg} which are comparable to the ones from BaBar in the region $m _{Z ^\prime} < 2 m _\mu$, but weaker than the BaBar-$4 \mu$ search in the $m _{Z ^\prime} > 2 m _\mu$ region (the bounds from BaBar and Belle II in the Fig. \ref{Plot} were adapted from Ref. \cite{Zhang:2020fiu}). \\

\paragraph{\label{APV}Parity violation}

A particularly powerful probe for the model with two doublets come from experiments sensitive to the weak charge $Q _W$ of protons and atomic nuclei.
A light $Z ^\prime$ axially coupled to the fermions introduces a new source of parity violation, which can manifest itself as a shift in $Q _W$ \cite{Bouchiat:2004sp, Campos:2017dgc, Arcadi:2019uif,Davoudiasl:2012qa}. The weak charge of the $^{133}Cs$ nucleus, obtained from precision measurements of the parity violating $6 S _{1/2} - 7 S _{1/2}$ atomic transition \cite{ParticleDataGroup:2020ssz},
\begin{equation}
Q _W ^{^{133}Cs, \text{exp}} = - 72.82(42) ,
\end{equation}
displays a slight discrepancy with the SM prediction $Q _W ^{^{133}Cs, \text{SM}} = - 73.16(35)$ \cite{Porsev:2009pr}, but is still consistent with the SM within $2 \sigma$.
For the proton, the weak charge can be measured in elastic electron-proton scattering with polarized electrons. The most recent measurement from the Qweak collaboration at JLAB \cite{Qweak:2018tjf} reads,
\begin{equation}
Q _W ^{p, \text{exp}} = 0.0719(45) ,
\end{equation}
in good agreement with the SM prediction $Q _W ^{p, \text{SM}} = 0.0711(2)$ \cite{ParticleDataGroup:2020ssz}. We show the bounds from these experiments in the Fig. \ref{Plot} as the light (dark) blue shaded region for the cesium (proton) weak charge. \\

\begin{figure*}[!t]
  \centering
  \includegraphics[width=0.48\linewidth]{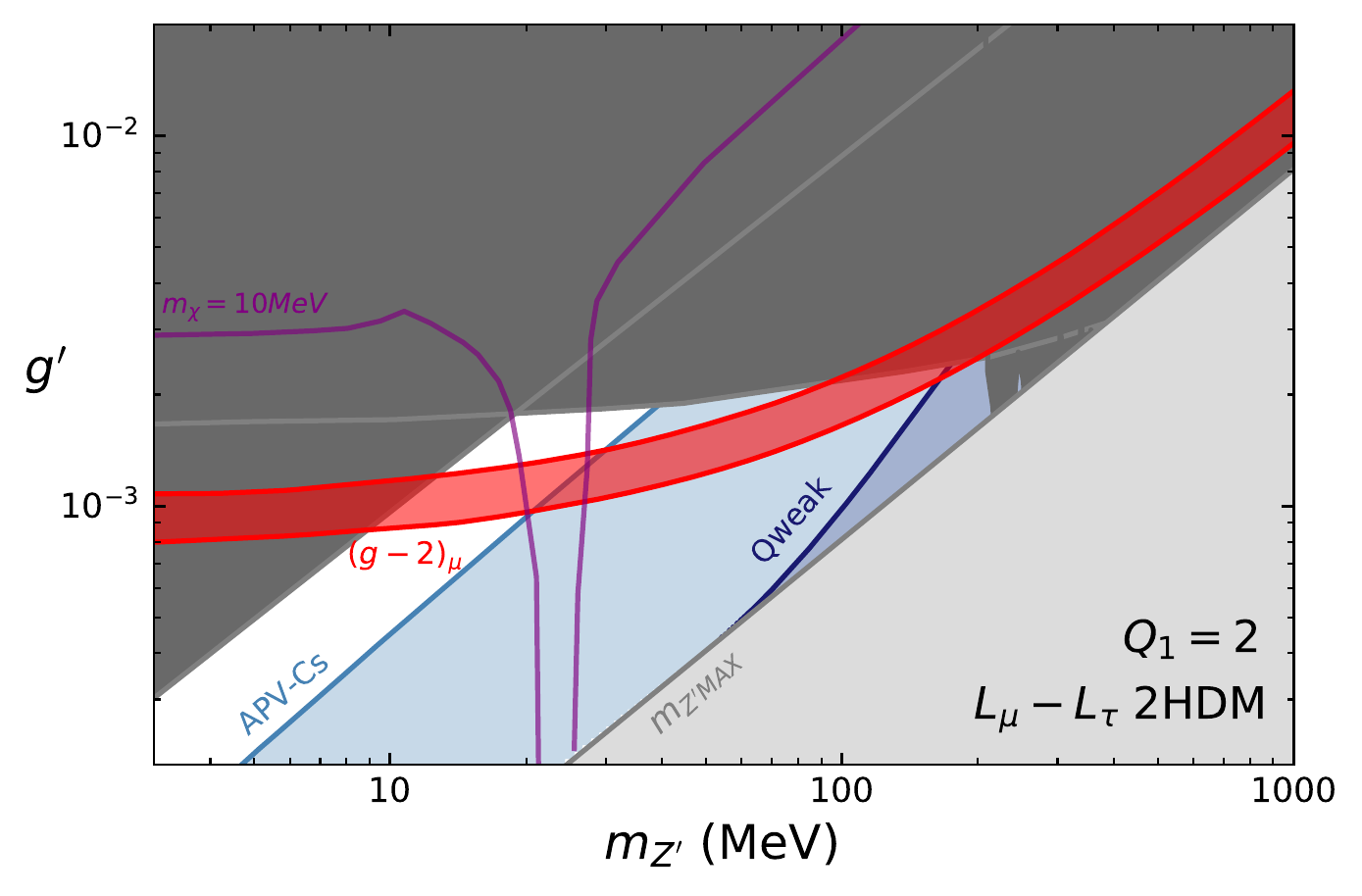}\quad
  \includegraphics[width=0.48\linewidth]{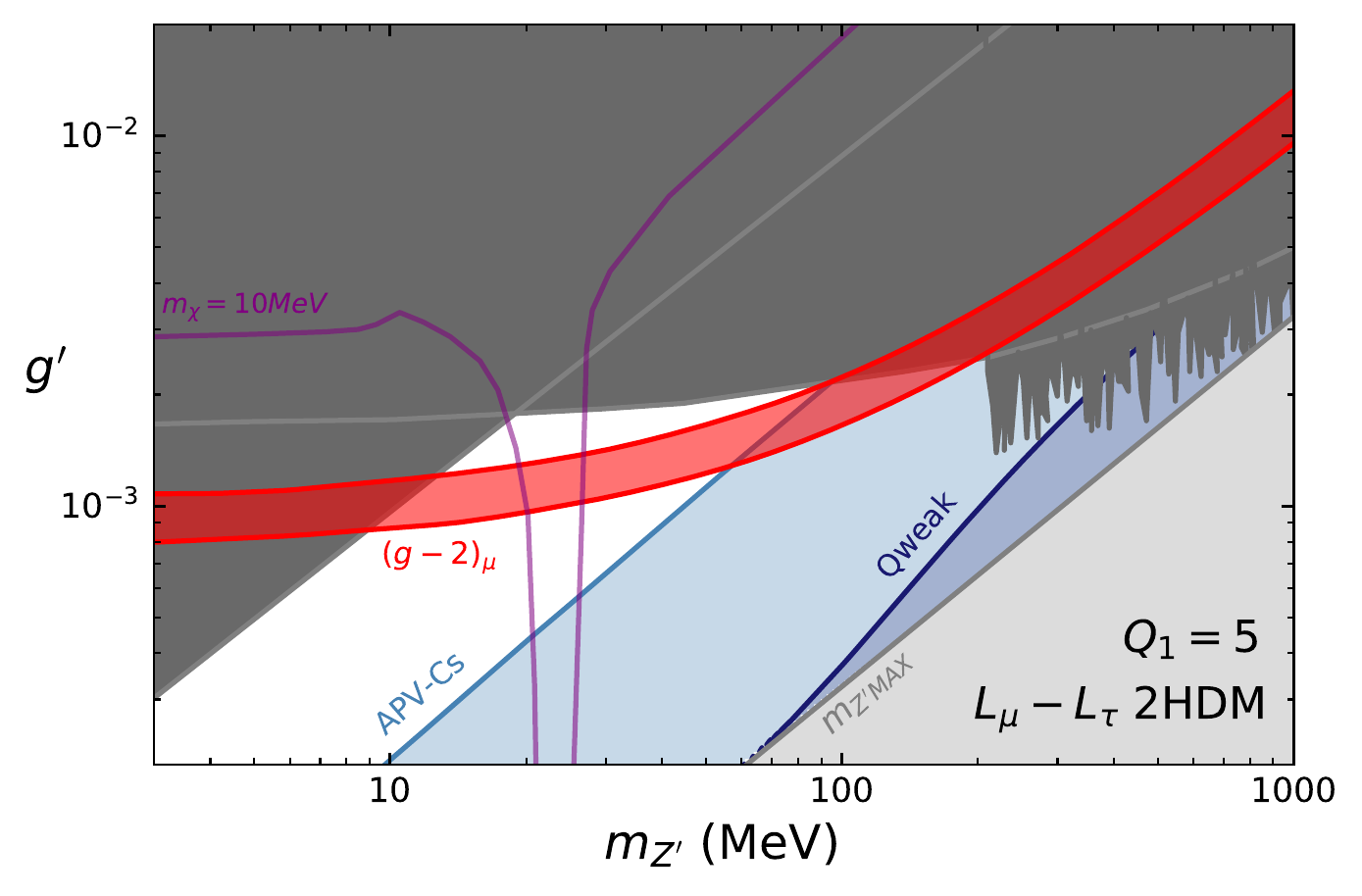}
  \caption{Effect of $\Phi _1$ doublet charge on the excluded parameter space. In the left (right) panel the charge of the $\Phi _1$ doublet is fixed as $Q _1 = 2$ ($Q _1 = 5$). The solid purple line show the correct dark matter relic abundance for a fixed dark matter $m _\chi = 10$ MeV.}
  \label{Plot2}
\end{figure*}

We see from Fig.\ref{Plot} that the several constraints from low energy experiments exclude most of the original parameter space suitable to accommodate the $(g-2) _\mu$ anomaly, except for the mass window between $10 \mbox{\ MeV} \lesssim m_{Z^\prime} \lesssim 200 \mbox{\ MeV}$ and $g ^\prime \sim 10 ^{-3}$. 
The $(g-2) _\mu$ band is sliced from above by the $4 \mu$ and neutrino trident searches at BaBar and CCFR, and from below by the Borexino neutrino-electron scattering data. 
It should be noted also the presence of an inaccessible region in the parameter space, which varies depending on the value of the $\Phi _1$ charge, depicted as the grey region in the lower right corner of Fig. \ref{Plot}. The line `$m _{Z ^\prime \text{\tiny{MAX}}}$' corresponds to the particular value $\tan \beta = 1$, in which $m _{Z ^\prime}$ reaches a maximum. Below this line, there is no $\tan \beta$ that fulfill the Eq. (\ref{zp_mass_eq}), which makes this a forbidden region. This is a particular feature that arises due to the symmetry being broken by a doublet, whose VEV is limited at the electroweak scale. The Fig. \ref{Plot2} shows how the forbidden region and the parity violation bounds vary with the doublet charge, where the left (right) panel corresponds to the $Q _1 = 2$ ($Q _1 = 5$) charge. It is apparent that higher $Q _1$ values lead to weaker constraints.

\section{Dark Matter}

Having the above $(g-2) _\mu$ parameter space in mind, could we also address dark matter?
In the context of dark matter particles with masses around the weak scale, the answer is no. Considering the thermal WIMP paradigm  a heavy dark matter requires a $Z^\prime$ which is too heavy to accommodate $(g-2) _\mu$. However, a light dark matter particle can, in principle, foot the bill. Although light dark matter accompanied by a light mediator is amenable to a wealth of constraints stemming from  BBN, ionization effects on the Cosmic Microwave Background, low energy accelerators, among others \cite{TheBABAR:2016rlg,Kaneta:2016uyt,Slatyer:2012yq,Nollett:2014lwa,Sirunyan:2018nnz}, the most relevant ones for the region of interest, $m_{Z^\prime} \simeq 10-200$~MeV, are displayed in Fig.\ref{Plot}. 

The simplest way to add dark matter in this model without spoiling gauge anomaly cancellation is by adding a vector like fermion $\chi$ with,

\begin{equation}
\mathcal{L} = \frac{g ^\prime}{2} \bar{\chi} \gamma ^\mu \chi Z _\mu ^\prime - m _\chi \bar{\chi} \chi ,
\end{equation}
where we implicitly adopted the $\chi$ charge under the $L_\mu-L_\tau$ symmetry to be equal to one. As the dark matter is very light, the s-channel mediated annihilations into SM particles will be the only relevant processes for the relic density. The other existing interactions are either kinematically prohibited or very suppressed, such as the processes mediated by the $Z$ boson that rises via $Z-Z^\prime$ mixing. 

After computing the dark matter relic density which is governed by the $Z^\prime$ interactions with SM fermions by solving the Boltzmann equation, we found that only near the $Z^\prime$ resonance we find a region of parameter space that sets the correct relic density without evoking non-standard cosmological histories, in other words, for $m_{\chi} \sim m_{Z^\prime}/2$. As the $Z^\prime$ mass should lie around $10-200$~MeV to explain $(g-2) _\mu$, we automatically know the dark matter mass as well, i.e. $m_{\chi} \simeq 5-100$~MeV. Quantitatively speaking, the correct relic density can be found at this mass range for $g^\prime\sim 10^{-3}$. In Fig. \ref{Plot2} we show a sample relic density curve for $m _\chi = 10$ MeV. One can easily see that this parameter overlaps with the favored region for $(g-2) _\mu$. Hence, with a light $Z^\prime$ and a vectorlike fermion we can nicely include a successful dark matter candidate in the scope of this 2HDM while explaining the $(g-2) _\mu$.

We emphasize that this setup relies on $Z^\prime$ resonance and thermal dark matter production. Although, one could certainly evoke non-standard cosmology, such as an entropy injection episode which would shift the relic density region, we will not dwell on the details of this mechanism as we can already reproduce the correct relic density within the thermal freeze-out scenario. The limits rising from direct and indirect dark matter detection are weaker than the ones shown in Fig. \ref{Plot2} in the region of interest, and for this reason are not shown.

\section{Discussion}

The several constraints from low energy experiments exclude most of the $(g-2) _\mu$ band, except for the mass range between $10 \mbox{\ MeV} \lesssim m_{Z^\prime} \lesssim 200 \mbox{\ MeV}$ and $g ^\prime \sim 10 ^{-3}$. In particular, the bounds from parity violation restrict the $(g-2) _\mu$ region within the 10-200 MeV window, which would otherwise remain unconstrained. Parity violation in this model is a direct consequence of the presence of the extra doublet, which induces axial $Z ^\prime$ couplings via the $Z ^\prime/Z$ mass mixing. Therefore the effect becomes stronger for larger $\epsilon _Z$, which occurs for high $Z ^\prime$  masses, close to the $m _{Z ^\prime \text{\tiny{MAX}}}$ line. For smaller $m _{Z ^\prime}$ (or equivalently larger $\tan \beta$, cf. Fig. \ref{zp_mass}), $\epsilon _Z$ becomes suppressed and the bound weakens. In the $0 < Q _1 < 5$ range, atomic parity violation from Cesium can exclude a large portion of the $(g-2) _\mu$ band unconstrained by other experiments, while for higher $Q _1$ values this bound becomes less severe. 

Most of the constraints discussed above apply in a similar manner to the $L _\mu - L _\tau$ singlet setup, as illustrated in the Fig. \ref{PlotSing1}. In comparison to the doublet case considered here, there is no upper limit on the $Z ^\prime$ mass and no parity violation bounds apply. Therefore, the lower right corner region excluded in the doublet case is allowed in the singlet scenario. In the case of a discovery of a $Z ^\prime$ signal associated with the $(g-2) _\mu$ in a future experiment, parity violation effects could be decisive in order to discriminate between the two scenarios, since these effects are a direct probe for the mass mixing generated in the presence of the doublet. Future experiments with polarized electron beams such as P2 at Mainz \cite{Becker:2018ggl} and Moller at JLAB \cite{MOLLER:2014iki} can extend the sensitivity to the $(g-2) _\mu$ region not reached by parity violation in Cesium. 

Another way to discriminate between the singlet and doublet cases is to directly search for the extra doublet scalars $H$ and $H ^+$ which, interestingly, can be lighter than the SM-like Higgs, as explained earlier. In fact, the authors of Ref. \cite{Arhrib:2017wmo} have found that in the type-I 2HDM, it is possible for $H$ to be as light as about 10 GeV and $H ^+$ to lie in the 80-160 GeV mass range without being in conflict with data from direct searches at LEP, Tevatron and LHC. It's been pointed out that in this scenario, provided that $m _{H ^+} < 175$ GeV, the $H ^+$ decay mode $H ^+ \rightarrow W ^+ H$ can dominate over the more traditional search channels $H ^+ \rightarrow \tau ^+ \nu$ and $H ^+ \rightarrow c s$, providing striking signatures for future LHC searches.

In a somewhat similar scenario to the one considered here, the authors of Ref. \cite{Lee:2013fda} have studied the extra Higgs decays in presence of a light $Z ^\prime$, obtaining good prospects for the detection of the $H$ boson through the process $pp \rightarrow H \rightarrow Z ^\prime Z ^\prime$, which leads to a clean final state with two pairs of collimated leptons. They also considered the detection of $H ^+$ in the channel $H ^+ \rightarrow W ^+ H$, with the final $H$ decaying to $H \rightarrow Z ^\prime Z ^\prime$ (differently from Ref. \cite{Arhrib:2017wmo}, in which $H \rightarrow \gamma \gamma$ is the dominant decay mode, since the $Z ^\prime$ is absent).
These search strategies can be adapted to the case of the 2HDM $L _\mu$-$L _\tau$ considered here, with possible improvements on the sensitivity, given that a larger $Z ^\prime$ branching fraction into muons can lead to better signal reconstruction in some cases. Although, in general, the results of these previous analysis apply qualitatively to the 2HDM $L _\mu$-$L _\tau$, it is clear that these collider signatures, and possibly novel ones, deserve a careful dedicated examination, which we defer to a future work.

Although not being the main focus here, a comment regarding neutrino masses is in order. Nonzero active neutrino masses can be generated straightforwardly in this model with the inclusion of right-handed neutrinos via type I seesaw mechanism. However, it is well known that the $L _\mu - L _\tau$ symmetry imposes a structure on the neutrino mass matrix and, despite being an excellent first order approximation for the observed pattern of neutrino mixing, an exact $L _\mu - L _\tau$ symmetry is now disfavoured by the data. Even though the $L _\mu - L _\tau$ symmetry is spontaneously broken here, the breaking using only a scalar doublet is in tension with current neutrino oscillation data when one considers both the measurements of the mixing angles obtained in oscillation experiments, and the Planck constraints on the sum of the active neutrino masses, so that the inclusion of extra scalars charged under $L _\mu - L _\tau$ might be necessary.

Furthermore, neutrino propagation through matter is affected by the new $Z ^\prime$ interactions, with potentially visible effects in oscillation experiments. General neutrino Non-Standard Interactions (NSI) are commonly parametrized by effective four-fermion operators $\nu _\alpha \nu _\beta \bar{f} f$, whose strengths relative to the Fermi constant are characterized by the dimensionless couplings $\epsilon _{\alpha \beta} ^f$, the so called NSI parameters, with $\alpha, \beta = e, \mu, \tau$ and $f = e, u, d$. The oscillation experiments are sensitive to the differences between the diagonal parameters $\epsilon _{\alpha \alpha} ^f - \epsilon _{\beta \beta} ^f$, hence non-universal neutrino interactions, as the ones in the $L _\mu - L _\tau$ model, can lead to observable effects. 
When combined with data from other neutrino experiments, such as $\nu$-e scattering or deep inelastic scattering, limits on individual $\epsilon$'s can be placed, the strongest ones currently being at the level of $ | \epsilon _{\alpha \alpha} ^f | \lesssim O(10 ^{-2})$ (see Ref. \cite{Farzan:2017xzy} for a recent review).

Constructing an explicit model with fully realistic neutrino oscillation parameters is beyond the scope of the present work (the interested reader can see, e.g., Refs. \cite{Xing:2015fdg, Asai:2018ocx, Zhou:2021vnf}). However, we emphasize that the incorporation of extra scalars can be done consistently within this framework without significantly change the phenomenology discussed above. Moreover, full agreement with the limits from the NSI parameters can be attained.

\begin{figure}[!t]
  \centering
  \includegraphics[width=\linewidth]{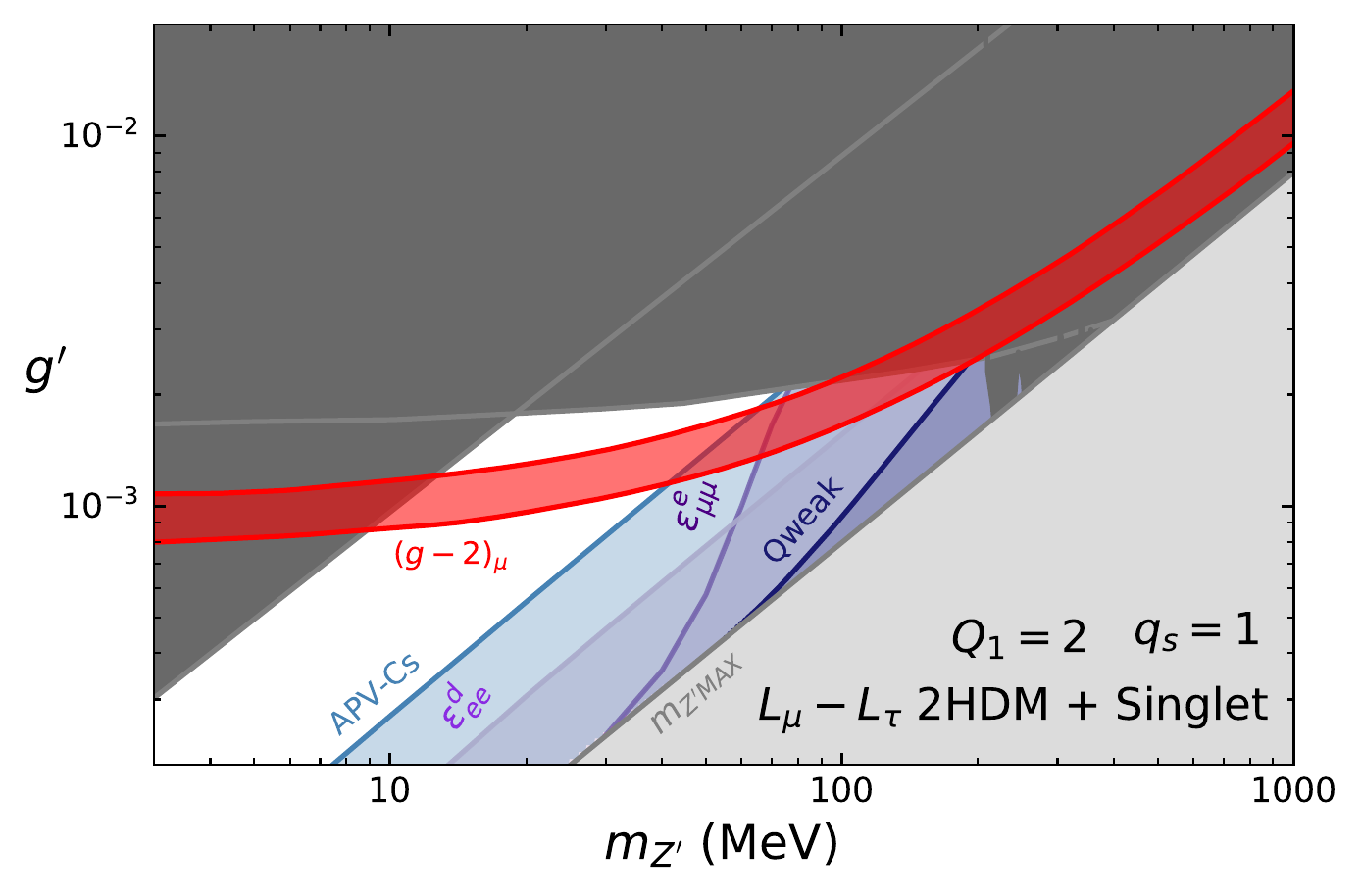}
  \caption{Effect of the inclusion of an extra singlet scalar of charge $q _s = 1$ and VEV $v _s = 50$ GeV to the model. The excluded region from parity violation is slightly modified in presence of the singlet in comparison to the doublet only case. The bounds from $\epsilon _{\mu \mu} ^e$ and $\epsilon _{ee} ^d$ NSI parameters are also shown.}
  \label{PlotSing2}
\end{figure}

We illustrate this point by assessing the effect of the addition of an extra scalar singlet to the model. Following the Ref. \cite{Asai:2018ocx} we assign a $L _\mu - L _\tau$ charge $q _s = 1$ for this scalar, which will give a contribution to the $Z ^\prime$ mass of order $g ^\prime q _s v _s$ through its vacuum expectation value $v _s$.
Thus, requiring that $v _s \lesssim 100$ GeV is sufficient to keep the $Z ^\prime$ mass in the right range for explaining the $(g-2) _\mu$. The extra contribution to $m _{Z ^\prime}$ will shift the grey forbidden region in Fig. \ref{Plot2} to the right, as now higher $Z ^\prime$ masses will be allowed. In addition, it will also indirectly modify the axial $Z ^\prime$ couplings to the fermions, since for a given $m _{Z ^\prime}$, the contribution from the singlet will enforce the doublet counterpart to be smaller, which amounts to an increasing on $\tan \beta$. Since a larger $\tan \beta$ yields a suppression on the mass mixing, the bounds from parity violation will be weakened. Such modifications are shown in Fig. \ref{PlotSing2} for $v _s = 50$ GeV, in which we can see no great departures from the previous results shown in Fig. \ref{Plot2}. 
In Fig. \ref{PlotSing2} it is also shown the bounds coming from the NSI parameters, in particular $ - 0.036 < \epsilon _{\mu \mu} ^e < 0.036 $ (95\% C.L.), coming from reactor and accelerator neutrino experiments \cite{Davidson:2003ha,Barranco:2007ej}, and $ -0.019 < \epsilon _{ee} ^d < 0.59 $ (95\% C.L.) \cite{Coloma:2017ncl}, from solar neutrino data combined with the recent observation of $\nu$-nucleus coherent scattering \cite{COHERENT:2017ipa}, which turn out to be the most relevant NSI contraints to the model.
As can be seen, these limits do not lead to further restrictions on the $(g-2) _\mu$ parameter space.

\section{\label{dc} conclusions}

After the observations of massive neutrinos and dark matter, arguably we have witnessed the first strong evidence for physics beyond the Standard Model, with the $(g-2) _\mu$ measurement at the Muon g-2 experiment. Motivated by this, we presented a Two Higgs Doublet Model featuring a light $Z^\prime$ within an $L_\mu-L_\tau$ symmetry. After discussing the relevant bounds, we have displayed the region of parameter space that explains $(g-2) _\mu$ in agreement with existing limits, highlighting the differences with respect to the usual $L_\mu-L_\tau$ scenario, where the $Z^\prime$ mass is generated by a singlet. Later, we discussed how a vectorlike dark matter candidate can be easily added into the model without spoiling gauge anomalies, while being able to reproduce the correct relic density. We pointed out that only near the $Z^\prime$ resonance, we can successfully have a light dark matter candidate while still having a $Z^\prime$ which is sufficiently light to explain $(g-2) _\mu$ and evade the aforementioned constraints. As the Muon g-2 has collected only 6\% of the data aimed by the collaboration, we seem to be on the verge to claim a groundbreaking discovery.

\acknowledgments

The authors thank Clarissa Siqueira for discussions. TM and FSQ thanks UFRN and MEC for the financial support. FSQ
is supported by the Sao Paulo Research Foundation (FAPESP) through grant $2015/158971$,
ICTP-SAIFR FAPESP grant $2016/01343-7$,
CNPq grants $303817/2018-6$ and $421952/2018-
0$, and the Serrapilheira Institute (grant number
$Serra-1912-31613$. YSV and ASJ acknowledge support from CAPES under the grants  $88882.375870/2019-01$ and $88887.497142/2020-00$. We thank the High Performance Computing Center (NPAD) at UFRN for providing computational resources.

\nocite{*}
\bibliography{ref}%

\end{document}